\pdfoutput=1
\documentclass[preprint,12pt]{elsarticle}
\usepackage{booktabs}
\usepackage{lineno,hyperref}
\ifpdf
\modulolinenumbers[1]
\usepackage{amsmath}
\usepackage{mathtools}
\usepackage{comment}
\usepackage{xcolor}
\journal{Journal of Medical Engineering \& Physics}

\begin{document}

\begin{frontmatter}

\title{A novel formulation for the study of the ascending aortic fluid dynamics with in vivo data}

\author[mymainaddress,mysecondaryaddress]{Katia Capellini}

\author[mymainaddress,mysecondaryaddress]{Emanuele Gasparotti}

\author[mythirdyaddress]{Ubaldo~Cella}
\author[myfourthaddress]{Emiliano~Costa}
\author[mymainaddress,mysecondaryaddress]{Benigno~Marco~Fanni}
\author[mythirdyaddress]{Corrado~Groth}
\author[mythirdyaddress]{Stefano~Porziani}
\author[mythirdyaddress]{Marco~Evangelos~Biancolini}
\author[mymainaddress]{Simona~Celi}\cortext[mycorrespondingauthor]{Corresponding author}
\ead{s.celi@ftgm.it}
\address[mymainaddress]{BioCardioLab, Fondazione Toscana Gabriele Monasterio, Massa, Italy}
\address[mysecondaryaddress]{Department of Information Engineering, University of Pisa, Pisa, Italy}
\address[mythirdyaddress]{Department of Enterprise Engineering, University of Rome Tor Vergata, Rome, Italy}
\address[myfourthaddress]{RINA Consulting SpA, Rome, Italy}

\begin{abstract}

Numerical simulations to evaluate thoracic aortic hemodynamics include a computational fluid dynamic (CFD) approach or fluid-structure interaction (FSI) approach. While CFD neglects the arterial deformation along the cardiac cycle by applying a rigid wall simplification, on the other side the FSI simulation requires a lot of assumptions for the material properties definition and high computational costs. The aim of this study is to investigate the feasibility of a new strategy, based on Radial Basis Functions (RBF) mesh morphing technique and transient simulations, able to introduce the patient-specific changes in aortic geometry during the cardiac cycle. Starting from medical images, aorta models at different phases of cardiac cycle were reconstructed and a transient shape deformation was obtained by proper activating incremental RBF solutions during the simulation process. The results, in terms of main hemodynamic parameters, were compared with two performed CFD simulations for the aortic model at minimum and maximum volume. Our implemented strategy copes the actual arterial variation during cardiac cycle with high accuracy, capturing the impact of geometrical variations on fluid dynamics, overcoming the complexity of a standard FSI approach. 

\end{abstract}

\end{frontmatter}

\section{Introduction}
The fluid dynamics of cardiovascular structures play a crucial role in the cardiovascular disease landscape \cite{Weigang2008,den2010hemodynamic,Qiao2011}. Quantification of the blood flow and associated bio-markers such as wall shear stress (WSS) and pressure gradients \cite{harloff2010} or specific hemodynamic indices can provide useful information in diagnosis and treatment of several cardiovascular pathologies.
There are two main techniques to quantify these hemodynamic parameters. The first is represented by \textit{in-vivo} time resolved phase-contrast magnetic resonance imaging (4D-PCMR/4D Flow MRI) scans. The second consists in patient specific \textit{in-silico} model using computational modelling.
Four-dimensional PCMR imaging made possible to non-invasively quantify patient-specific blood velocity profiles \cite{taylor2009patient, celi2017multimodality}, however, there are several limitations in both spatial and temporal resolution of the signals \cite{bakhshinejad2017, nayak2015}.

As a consequence of such limitations, in a quantitative way, hemodynamic indices are difficultly obtained or measured with a low accuracy \textit{in-vivo} \cite{Petersson2012}.
Numerical simulations by means of Computational Fluid Dynamics (CFD) and Fluid-Structure Interaction (FSI) turned out to be useful and effective tools to get an insight of vascular complex flows thanks to the possibility to explore patient-specific hemodynamics and to investigate different cardiovascular pathologies \cite{Caballero2013,fanni2020}. The combined use of high resolution computational models and data extracted from clinical imaging is increasingly becoming a reliable instrument to understand and predict the outcome of arterial disorders \cite{celi2014,zhong2018, celi2017multimodality}.
CFD simulations underwent a notable development in the cardiovascular field, representing an effective strategy to investigate the hemodynamics of thoracic aorta \cite{Bozzi2017,condemi2017}.
Although the rigid wall assumption adopted in the CFD approach could be a reasonable simplification in the cases of high arterial stiffness  \cite{Taylor2010,avolio2013}, this aspect is not negligible when considerable wall deformations occur during the cardiac cycle \cite{Vulliemoz2002}.
The rigid wall hypothesis has an important impact on the flow pattern evaluation due to the non consideration of wall compliance and geometry changes of the vessel, representing one of the most important determinant of local blood flow behaviour \cite{Jin2003,de2011,TSE2012}.
CFD disregards wall motion, e.g. the interaction between pulsatile blood flow and the compliant arterial wall, which may affect the estimation of WSS distribution at the aortic wall \cite{trachet2015, pons2020fluid}. 
In FSI models, the interaction of a deformable structure with the blood flow is computed \cite{Mendez2018,campobasso2018,boccadifuoco2018impact}, but it requires additional structural information such as material behaviour characterization as well as wall thickness and their spatial variations quantification, which are rather difficult to be defined \textit{in-vivo} \cite{Fanni_2020a, vignali2020modeling}.
In order to overcome this lack of information, a surrogate of vessel material behaviour can be retrieved from pulse wave velocity through the Moens–Korteweg equation \cite{avril2009vivo, Fanni_2020a}. 
Moreover, FSI simulation approach is a more complex tool and it needs consistent computational time and costs that currently made this technique difficult to insert in a clinical workflow. 
At present, imaging techniques are able to provide useful data in terms of wall motion and geometrical changes as well as in terms of fluid dynamic at specific phases of the cardiac cycle. 
Therefore, the development of a computational environment able to combine the wall motion of a vascular structure during the cardiac cycle and the CFD technique could be a promising alternative approach. This moving-boundary method (MBM) presents two main advantages: to provide a more accurate and realistic hemodynamic information while reducing the computational time and costs. 
The MBM is a well established technique \cite{groth2019fast, evangelos2019radial, papoutsis2019aerodynamic}, however it was investigated only by few authors in the cardiovascular landscape. Lantz et al.\cite{lantz2014} investigated the impact of wall motion on aorta hemodynamics by implementing a prescribed wall motion CFD simulation starting from 4D magnetic resonance (MR) images.\\
More recently, Capellini and co-workers (\cite{Porziani2017, Capellini2018, groth2018medical}) proposed a MBM approach integrating CT dataset, CFD simulations and Radial Basis Functions (RBF) mesh morphing technique. These studies were applied to investigate the fluid dynamic patterns in the ascending thoracic aortic aneurysm progressions and demonstrated the possibility to achieve a target geometry by covering also large shape modifications without losing accuracy in terms of mesh quality. RBF mesh morphing technique turned out to investigate the consequences of morphological variations on hemodynamics \cite{biancolini2012} or on the  positioning of a cannula \cite{gallo2014virtual}.

This work aims to develop a new fluid dynamic simulation strategy, based on the integration of RBF mesh morphing technique and transient CFD simulation, taking into account for the thoracic aorta deformation during cardiac cycle and thus overcoming the assumptions required by the FSI approach. The novel strategy feasibility and its accuracy were investigated together with the effects of geometry changes on the aortic hemodynamics and the results were compared with those from CFD assuming two geometries at different phases of the cardiac cycle.

\section{Radial Basis Function method}
RBF are a class of interpolation functions which focus on the recovery of unknown multivariable functions $f(\mathbf{x})$ from known data $(\mathbf{x})$ by computing an interpolator $s(\mathbf{x})$ expressed as a linear combination of basis functions $\phi$. In a general way, given a points set on a source surface $\Omega_S$, called source points ({$s_i$ with i=1,2,…,$N$}), and a points set on a target surface $\Omega_T$, called target points ({$t_j$ with j=1,2,…,$N$}), the problem consists in the determination of an optimal spatial interpolation functions $s(\mathbf{x})$ that deforms $\Omega_S$ to $\Omega_T$. The approximation of $f(\mathbf{x})$ is given by:

\begin{equation}\label{eq:interpolation}
f(\mathbf{x})\approx  s(\mathbf{x}) = \sum_{i=1}^{N}\alpha _i\phi_i(\left \|  \mathbf{x} -\mathbf{x_i}\right \|)
\end{equation}

where $\phi_i$ is a radial function depending on the Euclidean
distance between $\mathbf{x}$ and $\mathbf{x_i}$ and $\alpha _i$ is its associated coefficient.
In this work a Gaussian RBF was adopted for $\phi_i$ ($\phi(r)=e^{-r^2}$). 
The term RBF refers to a general formulation written as:

\begin{equation}\label{eq:interpolation1}
f(\mathbf{x})\approx  s(\mathbf{x}) = \sum_{i=1}^{N}\alpha _i\phi_i(\left \|  \mathbf{x} -\mathbf{x_i}\right \|) + h(\mathbf{x})
\end{equation}

where an additional polynomial part $h(\mathbf{x})$ is included to guarantee the existence and the uniqueness of the solution. The error of interpolation depends on the choice of RBF. Once the RBF function is defined, the RBF problem consists in the solution of the associated linear system.

\section{Material and Methods}
The work procedure was based on the integration of three main phases: the image processing (i), the RBF mesh morphing technique application (ii) and the execution of numerical simulations (iii). The feasibility together with the accuracy of the new implemented strategy to cope the actual aortic wall deformation during CFD transient simulation, hereafter called CFD$_{RBF}$, were tested. Additionally, CFD simulations were performed for the models with the largest and smallest aorta geometry in order to carry out a results comparison.  

\begin{figure}[h!]
    \centering
    \includegraphics[scale=0.41]{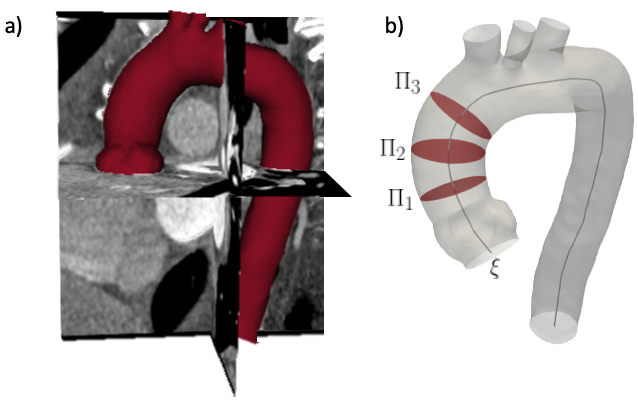}
    \caption{Example of images segmentation and of 3D reconstructed aorta model at phase 0$\%$ of the cardiac cycle (a); vessel centerline ($\xi$) and three cross sections ($\Pi_1, \Pi_2, \Pi_3$) (b).}
    \label{fig 1:segmentation}
\end{figure}

\paragraph{Image processing} The images were acquired with a 320-detector scanner (Toshiba Aquilon One, Toshiba, Japan), using iodinated contrast medium and presented an average pixel size of 0.468 mm and slice thickness of 0.5 mm. The acquisition protocols included a total body scan and 10 ECG-gated phases of the ascending aorta and aortic arch. All the CT datasets were retrospectively analysed. A segmentation process was performed by using a threshold algorithm with the same threshold level for each dataset (Figure \ref{fig 1:segmentation}a). 3D geometrical models of ascending aorta, arch and supra-aortic vessels were generated for each cardiac phase. The segmented models were exported as stereolithography (STL) file-format for accomplishing the mesh morphing process setup.
Based on the definition of the vessel centerline ($\xi$), defined as the geometrical locus of the centres of the maximal inscribed spheres in the vascular geometry, the aortic deformation was evaluated by calculating the sections diameters normal to $\xi$. Technically, the centerline was generated automatically by using VMTK on the 3D surface triangulation of each geometry.

\noindent The flow descriptors were evaluated for three cross sections shown in (Figure \ref{fig 1:segmentation}b): at the section characterized by maximum variation of diameter during the cardiac cycle ($\Pi_2$) and at two additional equally-spaced sections ($\Pi_1$ and $\Pi_3$).

\begin{figure}[h]
    \centering
    \includegraphics[scale=0.45]{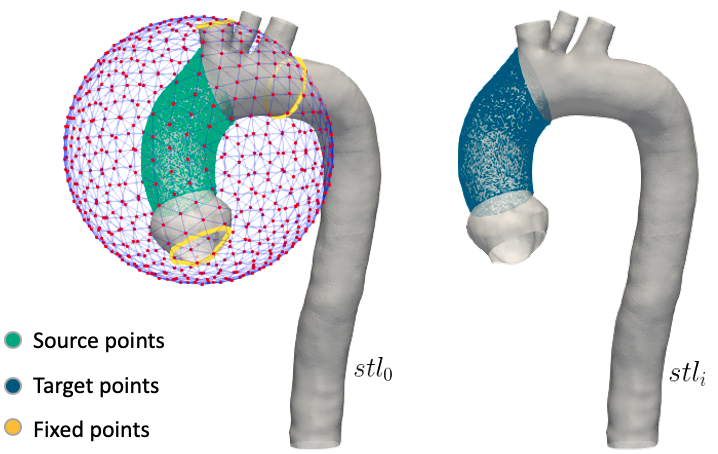}
    \caption{RBF mesh morphing setup: source points on $stl_0$ (a) and target points on $stl_i$ (b) points in green and blue respectively, fixed zone colored in yellow and sphere grid for the mesh morphing domain. For interpretation of the references to color in this figure legend, the reader is referred to the web version of this article.}
    \label{fig 2:RBF}
\end{figure}

\paragraph{RBF mesh morphing process}
In order to obtain a transient shape deformation of the ascending aorta geometry synchronised with the cardiac cycle, a specific simulation strategy was developed. This strategy is based on three main steps: calculation of the RBF solution targeting for each of the selected phases, the position of the aorta geometry with respect to its baseline configuration (i), calculation of the RBF incremental solution allowing to reshape the aorta over each interval of time (ii) and coupling the transient RBF solution with the CFD system (iii). Firstly, the RBF solutions of the selected phases were calculated by taking as reference the mesh of the aortic model at 0\% phase of cardiac cycle (baseline configuration) so that the associated displacement field from $stl_0$ to the $stl_i$ was obtained. To perform this, the source points were generated on the $stl_0$ geometry while specific sets of target points were selected on each $stl_i$ model (Figure \ref{fig 2:RBF}). Here, the morphing action was restricted to specific zones of the mesh detected with a sphere grid. The associated RBF solutions were referred to as: RBF$_{0-20}$, RBF$_{0-40}$, RBF$_{0-60}$ and RBF$_{0-80}$.

Secondly, the RBF solutions were then processed in order to obtain an incremental displacement field composed by several RBF solutions respectively named: RBF$_{0-20}$, RBF$_{20-40}$, RBF$_{40-60}$ and RBF$_{60-80}$. An additional RBF solution was implemented in order to handle the final part of the diastolic phase RBF$_{80-0}$, thus guaranteeing the closure of the complete cardiac cycle.  
Finally the incremental solutions were properly combined to automatically synchronize the actual shape with the pulsating fluid flow during CFD computing. To perform this strategy, RBF incremental solutions were activated with different time-dependent weights, by implementing a custom scheme in Fluent code together with the usage of RBF morph tool according to the following equation:

\begin{equation}\label{eq:RB}
R B F_{n, n+1}(t)=\left\{\begin{array}{lll}
0 & \text { if } \quad \mathrm{t} \leq t_{n} \\
\frac{t-t_{n}}{t_{n+1}-t_{n}} &  \text { if } \mathrm{t}_{n}<t<t_{n+1} \\
1 & \text { if } \quad \mathrm{t} \geq t_{n+1}
\end{array}\right.
\end{equation}

where the term $RBF_{n,n+1}$ represents the RBF solutions previously described at the second step (RBF$_{0-20}$, RBF$_{20-40}$, RBF$_{40-60}$, RBF$_{60-80}$ and RBF$_{80-0}$).
The whole process was implemented directly in Ansys\textsuperscript{\texttrademark} Fluent\textsuperscript{\texttrademark}by using the RBF Morph\textsuperscript{\texttrademark} add-on.
Two different kinds of evaluation were performed to assess the effectiveness of the RBF mesh morphing technique: one in terms of capability to reach the aorta geometry at each phase of deformation and another in terms of conservation of mesh quality. The first metric consists in calculating the distance between each 3D model reconstructed from images and corresponding RBF morphing configuration at that phase whilst the second evaluation was carried out by considering the skewness of the mesh at different stages of deformation.

\paragraph{Simulations setup} The governing equations of fluid motion, the Navier–Stokes equations, were numerically solved in their discrete form by applying the finite volume method. 
Three different CFD simulations were performed: a CFD simulations assuming the geometry at 0\% phase (CFD$_0$), a CFD assuming the geometry at 40\%  phase (CFD$_{40}$) of cardiac cycle and the new implemented transient simulation (CFD$_{RBF}$). 
For all the three CFD simulations, the same computational setup was assumed.
The 3D incompressible Navier-Stokes equation system used for blood flow simulation is given as:
\begin{equation}
\begin{matrix}
\rho \frac{\partial v}{\partial t} +\rho(v \cdot \nabla)v-\nabla\sigma = 0\\ 
\nabla \cdot v = 0
\end{matrix}
\end{equation}

where $\rho$ is the density of the fluid, $v$ is the blood velocity and $\sigma$ is the Cauchy stress. The blood was assumed as a non-Newtonian incompressible fluid (density of 1060 $kg/m^3$), adopting the Carreau-Yasuda viscosity model according to \cite{Avril2020} and described by Equation \ref{eq:viscosity}:

\begin{equation}\label{eq:viscosity}
\mu =\mu_{\infty}+(\mu_0-\mu_{\infty})\left [ 1+(\lambda\dot{\gamma})^{\alpha}) \right ]^{\frac{n-1}{\alpha}}
\end{equation}

where $\dot{\gamma}$ represents the scalar shear rate, $\mu_0$=0.042 Pa$\cdot$s is the blood viscosity at low shear rate, $\mu_{\infty}$=0.00345 Pa$\cdot$s is the blood viscosity at high shear rate, $\lambda$=3.313 s is the time constant, the power law index $n$=0.375 and Yasuda exponent $\alpha$=2$^{21}$. All models were discretized with a mesh of tetrahedral elements with an element size of 1 mm and prism elements were included by implementing a 4 inflation layers for a total thickness of 1.2 mm and a growth-rate equal to 1.5 to improve the accuracy in the WSS estimation. Regarding the boundary conditions, a blood flow velocity inlet profile was assigned to the aortic inlet and a pressure outlet condition was assumed for the four outlets: innominate artery (IA), left common carotid artery (LCCA), left subclavian artery (LSA) and descending aorta (DA). An idealised inlet flow previously adopted in \cite{Capellini2018} was used after a re-scale process in order to synchronise the flow curve with the cardiac cycle of the patient. The pressure profile along the cardiac cycle was obtained by implementing a lumped 3-element Windkessel model whose parameters, defined based on outlets areas distribution \cite{boccadifuoco2016uncertainty}, are reported in Table \ref{tab:RCR}. Three cardiac cycles were simulated and the results were evaluated at the last cycle in order to avoid the initial transient effects.

\begin{table}[!ht]
\small\sf\centering
\begin{tabular}{l|cccc}
\toprule
                           & IA          & LCCA        & LSA           & DA\\
\midrule
$R_d$   & $6.80e^{8}$ & $1.91e^{9}$ & $9.52e^{8}$   & $2.53e^{8}$      \\
$R_p$   & $4.34e^{7}$ & $1.22e^{7}$ & $6.08e^{7}$   & $1.62e^{7}$     \\
$C$  & $2.18e^{-9}$ & $7.74e^{-10}$ & $1.55e^{-9}$  &  $5.84e^{-9}$       \\
\bottomrule
\end{tabular}
  \vskip 0.2 cm
  \caption{Windkessel parameters: distal ($R_d$), proximal ($R_p$) resistances and compliance ($C$) for all outlets. Resistances are expressed in $kg~m^{-4}~s^{-1}$ while the compliances in $kg^{-1}~m^{4}~s^{2}$.}
  \label{tab:RCR}
\end{table}  

The hemodynamic results were investigated in terms of blood velocity over the three cross sections ($\Pi_1$, $\Pi_2$, $\Pi_3$) and by determining four hemodynamic indices in $\Pi_2$ and in the whole domain for all simulations.


\paragraph{Hemodynamic Indices} The differences in terms of flow eccentricity were assessed by calculating the normalized flow eccentricity (NFE) \cite{sigovan2011}, defined as the Euclidean distance between the vessel centre and the centre of velocity ($C_{vel}$) for the forward flow in a selected section of the aorta, and normalized to the lumen radius. $C_{vel}$ was calculated as the weighted centroid of the section. Each mesh node was weighted by the corresponding velocity value as defined in the following equation, where $r$ is the radius and $V$ the velocity is computed as:
\begin{equation}\label{eq:1}
    C_{vel} = \frac{\sum_{i}r_{ij}|V_i|}{\sum_{i}|V_i|} \quad j=x,y,z \quad i=\textrm{mesh section vertices}
\end{equation}

A NFE value equal to 1 implies that the flow has its maximal eccentricity, whereas NFE equal to 0 indicates a central distribution of flow.
The helical structures of blood flow were also evaluated in terms of localized normalized helicity (LNH) that describes the cosine of the angle between the vorticity and velocity vectors as proposed in \textcolor{blue}{\cite{shtilman1985,morbiducci2009vivo}}:
\begin{equation}\label{eq:2}
    \mbox{LNH}(s,t) = \frac{V(s,t)\cdot\omega(s,t)}{|V(s,t)||\omega(s,t)|} 
\end{equation}
where $s$ is the position, $t$ the time instant, $V$ is the velocity vector and ~$\omega$ the vorticity vector.
The LNH index ranges from -1 to +1, denoting a left-handed (counter-clockwise) and a right-handed (clockwise) rotation respectively \cite{morbiducci2007helical}. High values of LNH reflect a completely helical flow while LNH close to zero indicates symmetric flow patterns. The helicity was also evaluated by index $h_1$ (time-averaged value of the helicity normalized with respect to volume), index $h_2$ (helicity intensity) and index $h_3$ (ranging between -1 and 1: it is positive (negative) when left-handed (right-handed) helical structures are predominant in the volume) \cite{gallo2012}.
These indexes are defined by the following equations:
\begin{equation}\label{eq:h index1}
    \mbox{$h_1$} = \frac{1}{TV} \int_{T}\int_{V}H_{k}dV dt 
\end{equation}
\begin{equation}\label{eq:h index2}
    \mbox{$h_2$} = \frac{1}{TV} \int_{T}\int_{V}|H_{k}|dV dt 
\end{equation}
\begin{equation}\label{eq:h index3}
    \mbox{$h_3$} = \frac{h_1}{h_2} 
\end{equation}
where T is the interval of integration corresponding to the period of cardiac cycle, V is the entire volume of domain and $H_{k}$ is the helicity density.
The WSS analysis included the evaluation of time-averaged wall shear stress (TAWSS) \cite{malek1999} and the oscillatory shear index (OSI) \cite{fytanidis2014}. The TAWSS parameter was calculated by integrating WSS magnitude of each mesh vertex over the cardiac cycle as described by the following equation:
\begin{equation}\label{eq:3}
    \mbox{TAWSS} = \frac{1}{T} \int_{0}^{T} |WSS(s,t)| \cdot dt
\end{equation}
The OSI index identifies the presence of high oscillatory direction of WSS during the cardiac cycle and it was calculated according to:
\begin{equation}\label{eq:4}
   \mbox{OSI} = 0.5\left[1-\left(\frac{|\int_{0}^{T} WSS(s,t) \cdot dt|}{\int_{0}^{T} |WSS(s,t)| \cdot dt}\right)\right]
\end{equation}
According to previous studies \cite{lee2008geometry,morbiducci2013inflow}, threshold values were specified for both TAWSS and OSI indices respect to the distributions on the ascending aorta for CFD$_{0}$ case. The threshold values of 10th and 20th percentiles were considered for TAWSS, while threshold values of 80th and 90th percentiles were calculated for OSI. For this last index, the surface area characterized by values beyond (below for TAWSS) the percentile values was calculated to better quantify disturbed shear.
 
\section{Results}
\begin{figure}[h!]
    \centering
    \includegraphics[scale=0.32]{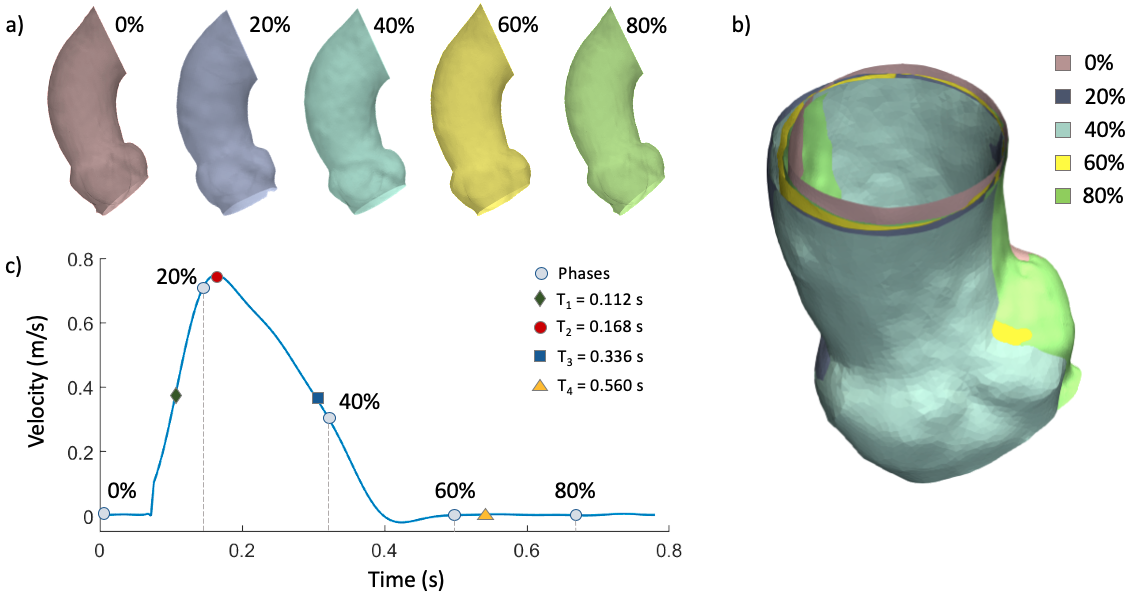}
    \caption{3D reconstructed geometries for five phases of cardiac cycle (a); example of overlay of the five models (b);
    flow rate waveform imposed at the inlet
section with the cardiac phases used in the CFD$_{RBF}$ procedure (0$\%$, $20\%$ $40\%$ $60\%$ $80\%$). The following characteristic time instants are highlighted: the maximum acceleration ($T_1$), the peak systole ($T_2$), the maximum deceleration ($T_3$) and the early diastole ($T_4$) (c). }
    \label{fig:segmentate}
\end{figure}

\paragraph{Image processing} The reconstruction of 3D models was successfully carried out for all cardiac phases. The evaluation of different geometries showed that the main variation along the cardiac cycle was in the ascending aorta, while no significant deformation was observed in the aortic arch. The maximum cross section diameter along the ascending aorta ranged between 27.9 mm (at 0\% phase) and 30.2 mm (at 40\% phase). The comparative analysis of each consecutive 3D model revealed very low variations. As a consequence of that, only geometries at intervals of 20\% were used to develop our mesh morphing process (Figure \ref{fig:segmentate}a). In Figure \ref{fig:segmentate}b the portion of the five ascending regions are overlapped. Figure \ref{fig:segmentate}c shows the aortic blood velocity profile with the five phases.

\begin{figure}[h!]
    \centering
    \includegraphics[scale=0.63]{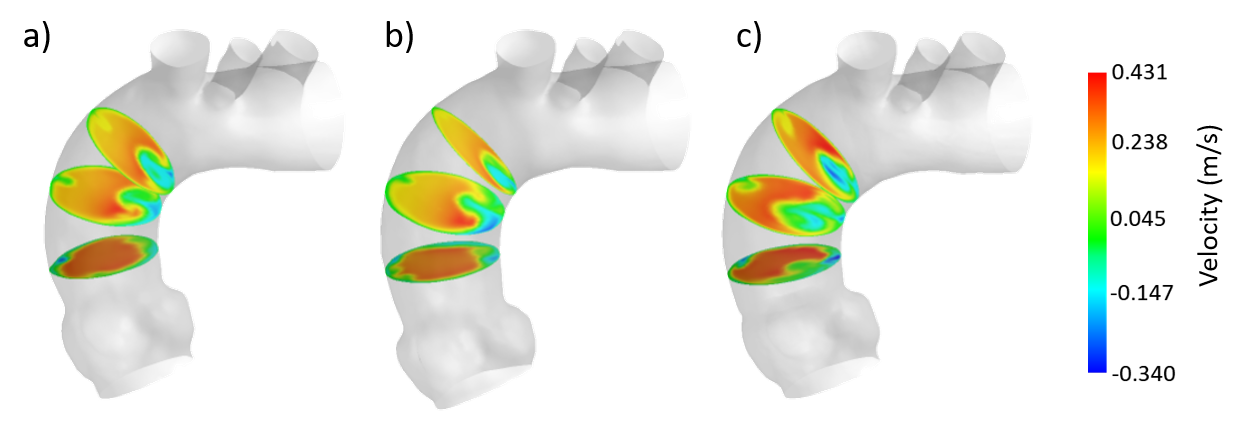}
    \caption{Through-plane velocity at three selected sections of ascending aorta at $T_3$ time of cardiac cycle for each performed simulation: CFD$_0$ (a), CFD$_{40}$ (b) and CFD$_{RBF}$ (c).}
    \label{fig:piani}
\end{figure}

\paragraph{Numerical Simulations} The comparison of the different simulations strategies in terms of hemodynamic parameters evaluation are presented at four different times of cardiac cycle to cope the behaviour at the maximum acceleration ($T_1$=0.112 s), at the peak systole ($T_2$=0.168 s), at the maximum deceleration ($T_3$=0.336 s) and at the early diastole ($T_4$=0.56 s) (Figure \ref{fig:segmentate}c). Figure \ref{fig:piani} depicts the through-plane velocity distributions in $\Pi_1$, $\Pi_2$ and $\Pi_3$ for the three numerical simulations at $T_3$.

\begin{figure}[h!]
    \centering
    \includegraphics[scale=0.66]{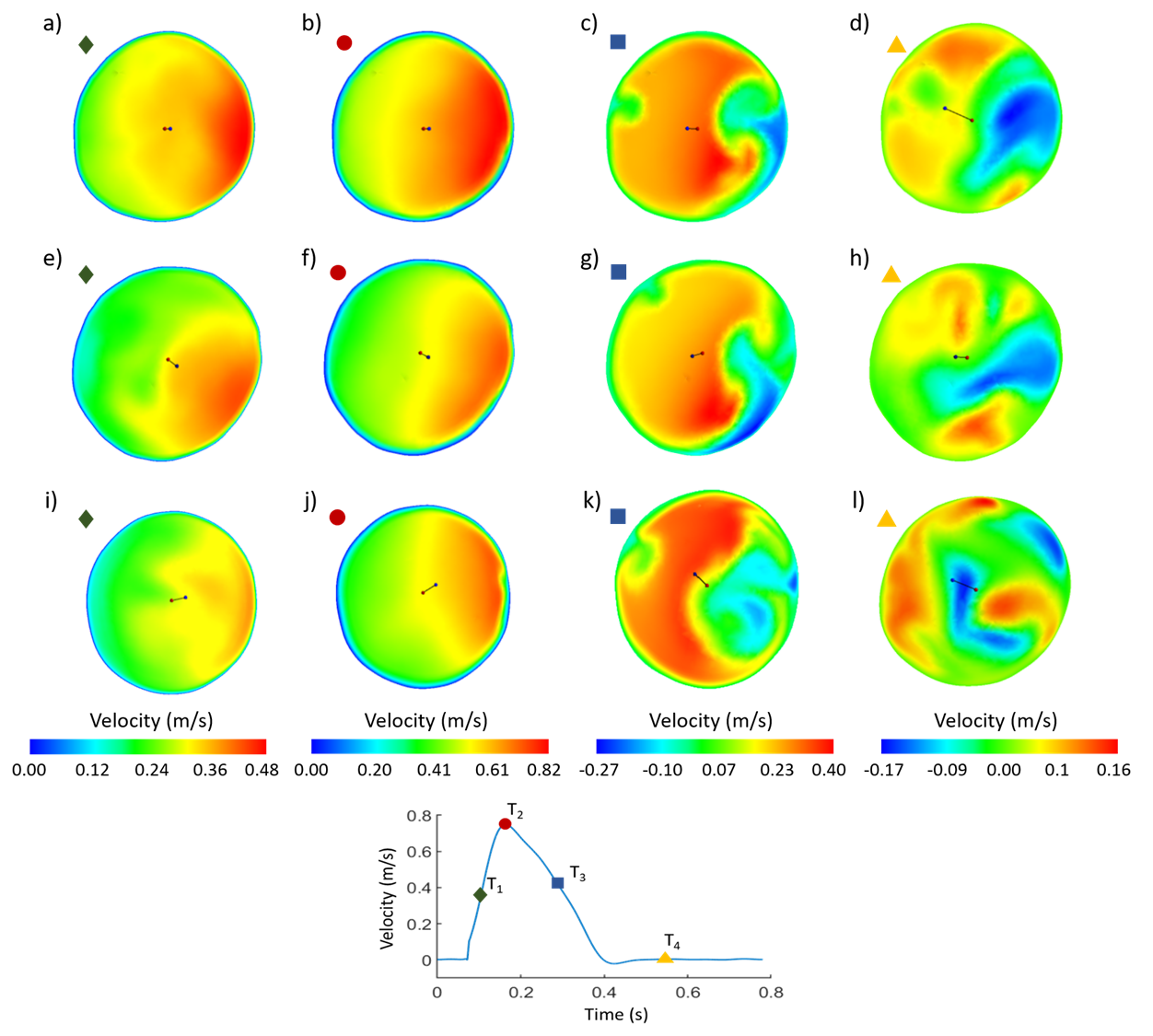}
    \caption{Plot of the $C_{vel}$ (in blue) and the centroid (in red) of the cross section at four different times of cardiac cycle ($T_1$, $T_2$, $T_3$ and $T_4$) for each performed simulation: CFD$_0$ (a-d), CFD$_{40}$ (e-h) and CFD$_{RBF}$ (i-l).}
    \label{fig:NFE}
\end{figure}

The distance between $C_{vel}$ and the centroid of the sections $\Pi_2$ together with the through-plane velocity distribution are depicted in Figure \ref{fig:NFE} in order to evaluate also the displacement of velocity centre. The NFE index values calculated on the same $\Pi_2$ plane are reported in Table \ref{tab:NFE}.

\begin{table}[h!]
\small\sf\centering
\begin{tabular}{l|cccc}
\toprule
             & $T_1$   & $T_2$   & $T_3$  & $T_4$\\
\midrule
CFD$_0$      & $0.06$  & $0.06$  & $0.11$ & $0.32$\\
CFD$_{40}$     & $0.12$  & $0.09$  & $0.11$ & $0.12$\\
CFD$_{RBF}$  & $0.161$ & $0.17$  & $0.18$ & $0.27$\\
\bottomrule
\end{tabular}
  \vskip 0.2 cm
  \caption{NFE values at $\Pi_2$ section for all simulations CFD$_0$, CFD$_{40}$, CFD$_{RBF}$ at the four selected instants of the cardiac cycle.}
  \label{tab:NFE}
\end{table} 

\begin{figure}[b!]
    \centering
    \includegraphics[scale=0.76]{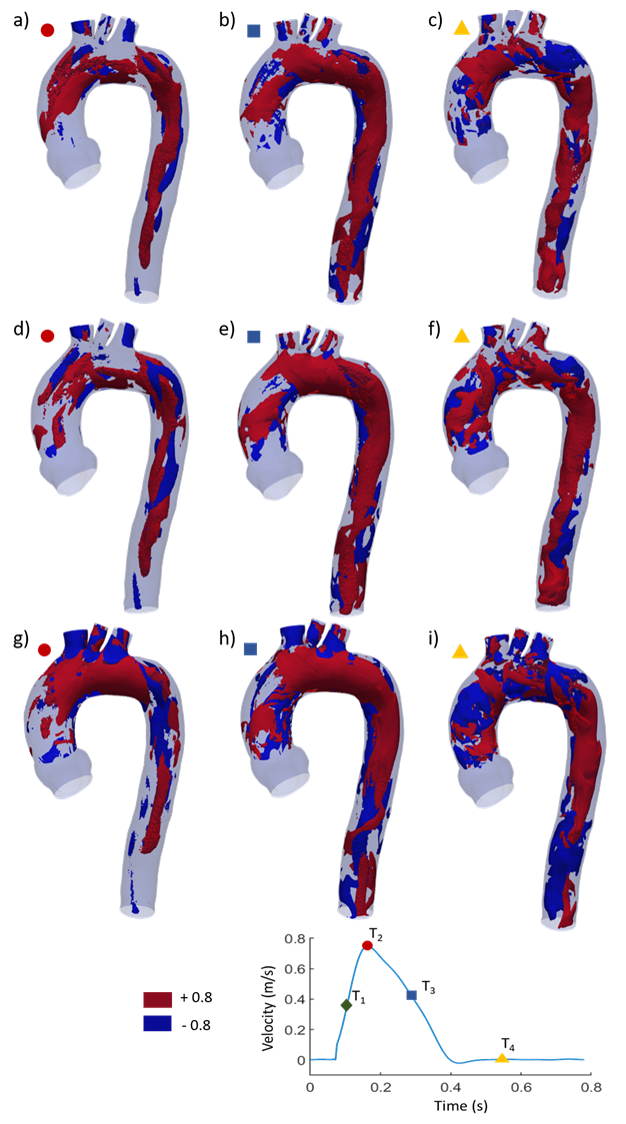}
    \caption{Isosurfaces of LNH at three different times of cardiac cycle ($T_2$, $T_3$ and $T_4$) for each performed simulation: CFD$_{0}$ (a-c), CFD$_{40}$ (d-f) and CFD$_{RBF}$ (g-i). High threshold values ($\pm$0.8) are adopted to visualize the markedly helical structures. For interpretation of the references to color in this figure legend, the reader is referred to the web version of this article.}
    \label{fig:LNH}
\end{figure}

The distribution of the LNH values in terms of isosurfaces with a threshold of $\pm$0.8 at $T_2$, $T_3$ and $T_4$ is reported in Figure \ref{fig:LNH} in order to visualize the markedly helical flow structures in the flow depending on the simulation strategy adopted. The presence of two types of helical flow was revealed: left handed rotation (blue color) and right-handed (red color). At systolic peak ($T_2$) a predominance of right-handed rotation is notable for the CFD$_0$ in the ascending region (Figure \ref{fig:LNH}a). This behaviour is less marked for the CFD$_{40}$ (Figure \ref{fig:LNH}d).
At the same timestep, the CFD$_{RBF}$ presents significant fluid structures with both rotation directions (higher LNH magnitude value) in the arch region (Figure \ref{fig:LNH}g). At $T_3$ CFD$_{40}$ and CFD$_{RBF}$ simulations present similar high LNH flow patterns (Figure \ref{fig:LNH}e,h) and lower values characterized CFD$_0$ in the arch region (Figure \ref{fig:LNH}b). With respect to the CFD solutions, the CFD$_{RBF}$ appears to be more effective for the supra-aortic vessels where a left-handed rotation is more pronounced. At $T_4$, the CFD simulations (Figure \ref{fig:LNH}c,f) show a decrease of helicity of the flow along all the aortic vessel, while the CFD$_{RBF}$ is characterised by a more pronounced left-handed rotation component in particular in the ascending region (Figure \ref{fig:LNH}i). The discrepancy in terms of helical flow between the CFD and the CFD$_{RBF}$ models is also revealed by the quantitative helicity indices. Even if the helicity intensity index ($h_2$) does not present significant differences (3.314, 3.022 and 3.289 for CFD$_0$, CFD$_{40}$ and CFD$_{RBF}$ respectively), the predominant sense of rotation stresses out the different behaviour of the  CFD$_{RBF}$ with respect to the CFD models.
\noindent 
In fact, $h_1$ and $h_3$ indices revealed a significant difference in terms of sign between the two approaches: positive for both CFD$_0$ (0.077 and 0.023) and CFD$_{40}$ (0.258 and 0.085) and negative for the CFD$_{RBF}$ (-~0.015 and -~0.028).

\begin{figure}[h!]
    \centering
    \includegraphics[scale=0.7]{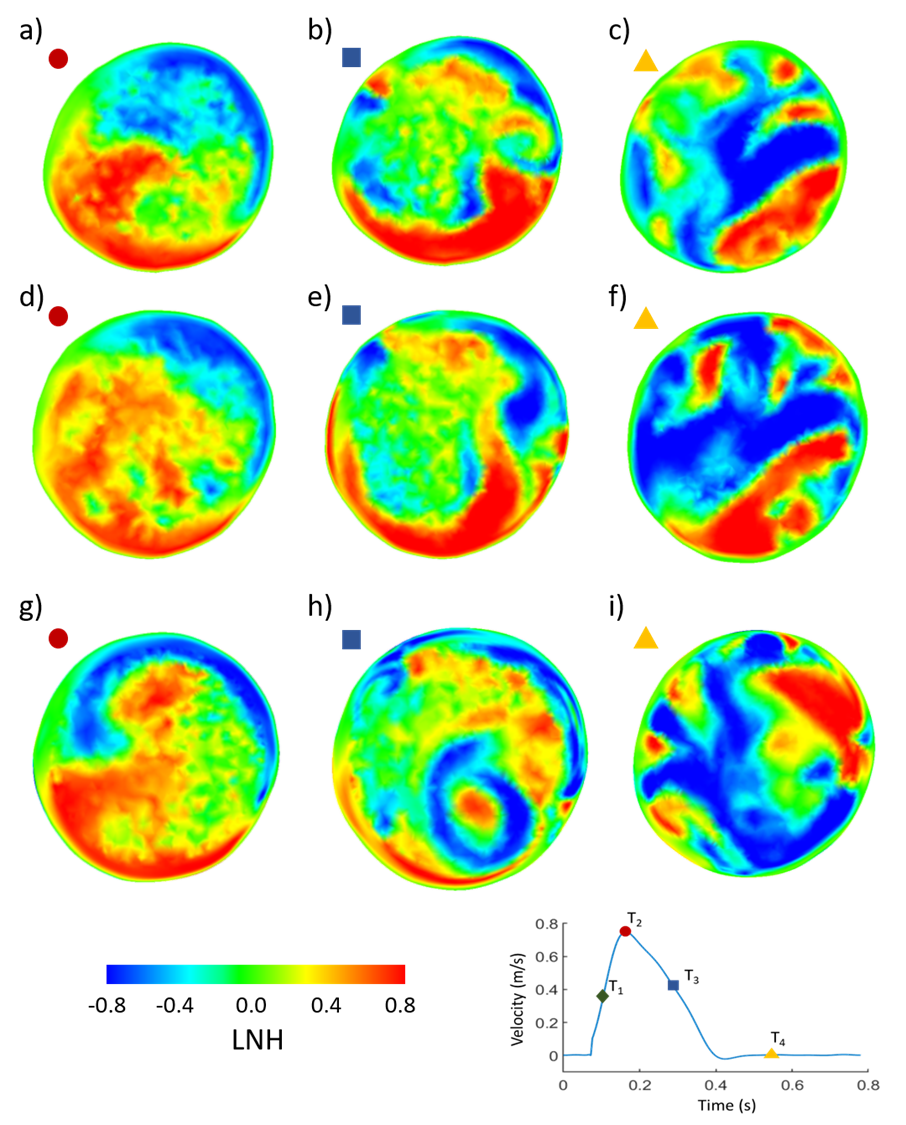}
    \caption{Projection on $\Pi_2$ of the LNH at three different times of cardiac cycle ($T_2$, $T_3$ and $T_4$) for each performed simulation: CFD$_{0}$  (a-c), CFD$_{40}$ (d-f) and CFD$_{RBF}$ (g-i).}
    \label{fig:LNH_in_plane}
\end{figure}

\newpage
\noindent Figure \ref{fig:LNH_in_plane} stresses out the map of the projection of LNH  on $\Pi_2$ at three different times of cardiac cycle ($T_2$, $T_3$ and $T_4$).

\begin{figure}[h!]
    \centering
    \includegraphics[scale=0.36]{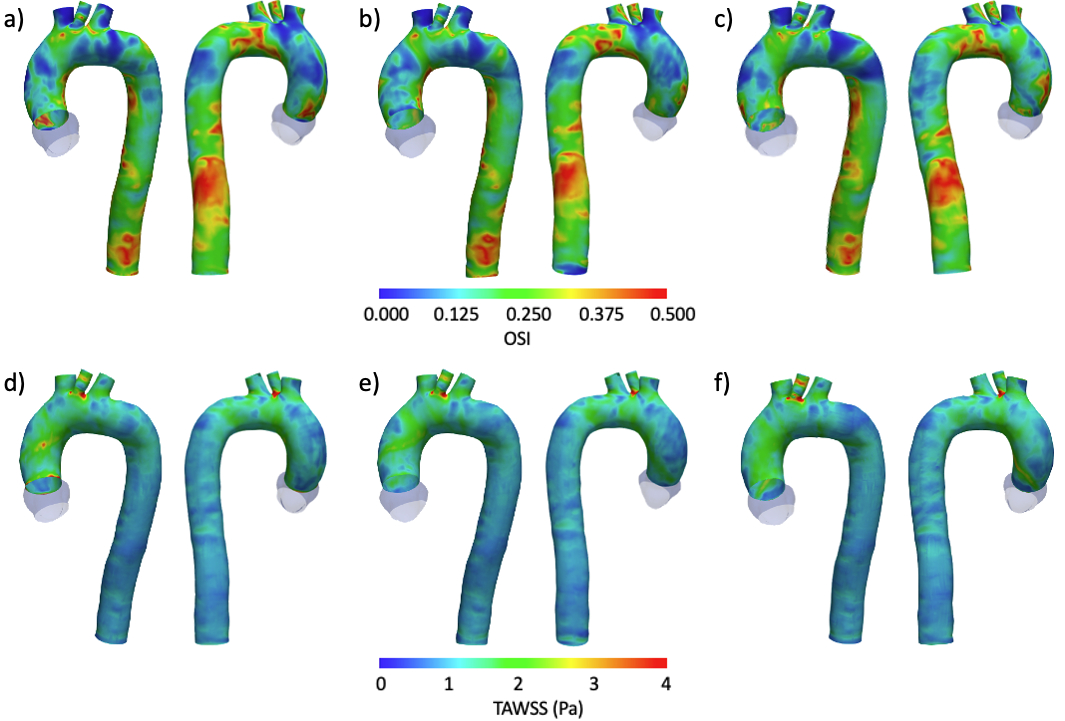}
    \caption{OSI and TAWSS distributions for for each performed simulation: CFD$_0$ (a,d), CFD$_{40}$ (b,e) and CFD$_{RBF}$ (c,f) respectively.}
    \label{fig:TAWSS_OSI}
\end{figure}

The OSI maps are shown in Figure  \ref{fig:TAWSS_OSI}(a-c). 
It is possible to notice different distributions of the two WSS indicators among the two simulation approaches. As expected, the
CFD$_{0}$ and CFD$_{40}$ depict differences in the ascending region but very similar maps in the remaining aorta. A higher OSI value in a wider ascending area for the CFD$_{RBF}$ with respect CFD simulations can be observed. This last observation is confirmed also by the TAWSS map distribution (Figure \ref{fig:TAWSS_OSI}(d-f)).

\begin{figure}[h!]
    \centering
    \includegraphics[scale=0.38]{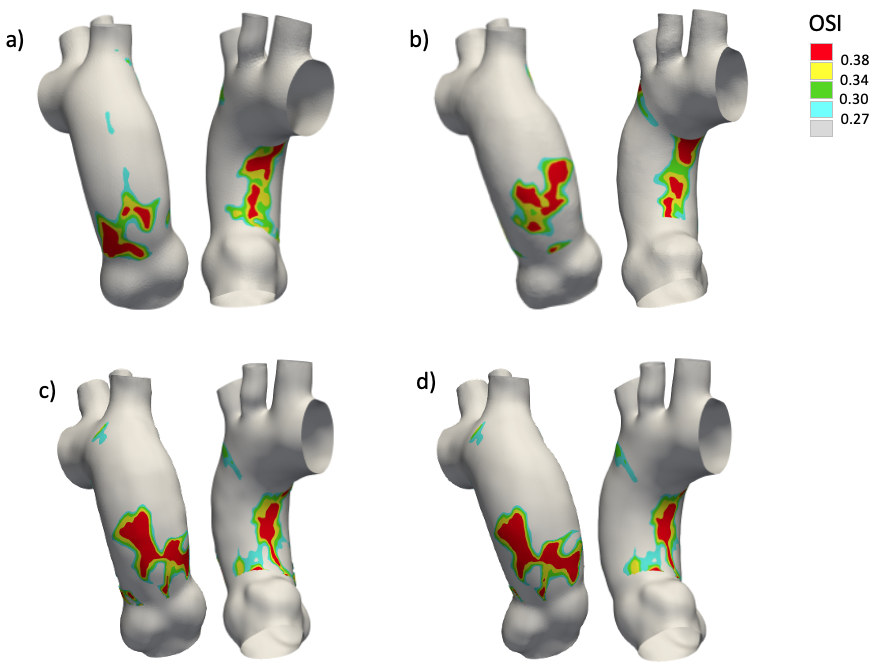}
    \caption{OSI distribution beyond threshold values in ascending aorta for CFD$_0$ (a), CFD$_{40}$ (b), CFD$_{RBF}$ at phase 0 of the cardiac cycle (c) and CFD$_{RBF}$ at phase $40$ of the cardiac cycle (d). Contour levels correspond to the 80th, 85th, 90th and 95th of the percentile values.}    \label{fig:MapsPercentileOSI}
\end{figure}

\begin{figure}[h!]
    \centering
    \includegraphics[scale=0.43]{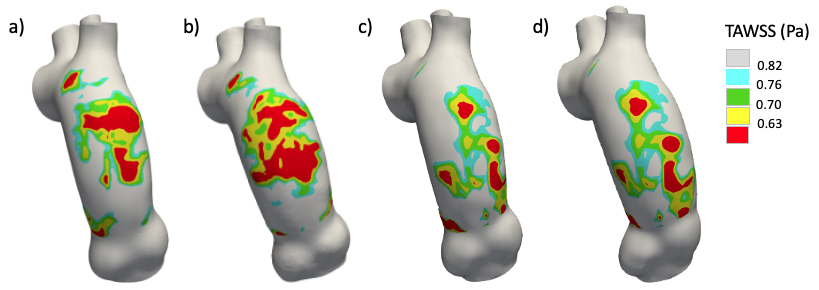}
    \caption{TAWSS distribution below threshold values in ascending aorta for CFD$_0$ (a), CFD$_{40}$ (b), CFD$_{RBF}$ at phase 0 of the cardiac cycle (c) and CFD$_{RBF}$ at phase $40$ of the cardiac cycle (d). Contour levels correspond to the 80th, 85th, 90th and 95th of the percentile values.} 
    \label{fig:MapsPercentileTAWSS}
\end{figure}

\noindent With particular attention to the ascending region, Table \ref{tab:Percentile} reports the percentage of  area with OSI (beyond) and TAWSS (below) values for the prescribed thresholds.\\
\noindent To allow the comparison process, morphological configurations at phases 0 and 40 of the cardiac cycle were reported for the CFD$_{RBF}$ case. The maps of surface area of ascending aorta with values beyond/below percentiles of OSI and TAWSS are shown in Figure \ref{fig:MapsPercentileOSI} and Figure \ref{fig:MapsPercentileTAWSS} respectively. The CFD$_{RBF}$ cases show a significant different topology with respect to CFD simulations: a larger area for critical OSI values and a smaller region for critical TAWSS.

\begin{table}[h!]
\small\sf\centering
\begin{tabular}{l|cccc}
\toprule
            & OSI (80th)  & OSI (90th) & TAWSS (20th)   & TAWSS (10th)   \\
\midrule
CFD$_0$     & $19.21\%$ & $9.91\%$ & $21.12\%$  & $10.74\%$  \\
CFD$_{40}$   & $21.00\%$ & $10.94\%$  & $26.92\%$  & $16.73\%$  \\
CFD$_{RBF}$ (phase 0) & $23.56\%$ & $11.81\%$ & $20.06\%$ & $9.49\%$  \\
CFD$_{RBF}$ (phase 40) & $23.50\%$ & $11.78\%$ & $21.36\%$ & $9.87\%$  \\
\bottomrule
\end{tabular}
  \vskip 0.2 cm
  \caption{Surface area of ascending aorta exposed to indices below TAWSS and beyond OSI for CFD$_0$, CFD$_{40}$, CFD$_{RBF}$. The percentage are related to the relative ascending area.}
  \label{tab:Percentile}
\end{table}

\newpage
\section{Discussion and conclusions}

In this study, we exploited the appealing concept of obtaining accurate hemodynamic solutions using a moving-boundary method approach and thus without invoking material descriptors that are difficult to retrieve. Our MBM is based on RBF mesh morphing techniques with radial basis function to obtain a subject-specific aortic model starting from a STL geometry derived from the segmentation of CT scan data.
In computational hemodynamics, simulations are commonly based on CFD approaches \cite{leuprecht2002combined, romarowski2018patient, zhu2018} thanks to their main advantages such as a reduced computational time and cost as well as the need of a more simple setup. Alternative strategies based on FSI were applied in literature \cite{reymond2013,baumler2020, pons2020fluid}. 
The influence of geometry on fluid dynamic patterns is recognized to be not negligible and the aortic wall changes during the cardiac cycle need to be considered \cite{jin2003effects, singh2016effects}. With this in mind, FSI simulations are more appealing because able to cope the physical problem in a more realistic manner. However, FSI simulations are more complex to setup and the assumption that they require can influence the entire numerical solution \cite{celi2012biomechanics, alimohammadi2015aortic, boccadifuoco2018}. For this reason, the incorporation of fully personalized, \textit{in-vivo} measurements of wall dynamics could be an alternative step in the pipeline for subject-specific analysis of the aortic hemodynamics.
First of all, this work revealed how the choice of different image phases for the model reconstruction to be used in standard CFD simulations also involves differences in hemodynamic results. Hence, the rigid wall simulations include an additional source of uncertainty due to the different possible geometries that can be used for the same subject.
Moreover, here we observe that: when a tomographic dataset is available, the application of our transient CFD$_{RBF}$ strategy is feasible; the application of a mesh morphing approach leads to the maximum distance error bounded below 0.24 mm and to a mesh distortion less then 0.035\% (for cell with skewness values \textgreater0.8 located in the morphing domain).
Our findings also show differences in WSS-based descriptors between CFD and CFD$_{RBF}$.
This result highlighted the relationship between wall motion and hemodynamics and it is in agreement with previous studies that showed that CFD simulation underestimated WSS on average by 10–30\% compared to FSI \cite{chandra2013fluid}.

\begin{figure}
    \centering
    \includegraphics[scale=0.38]{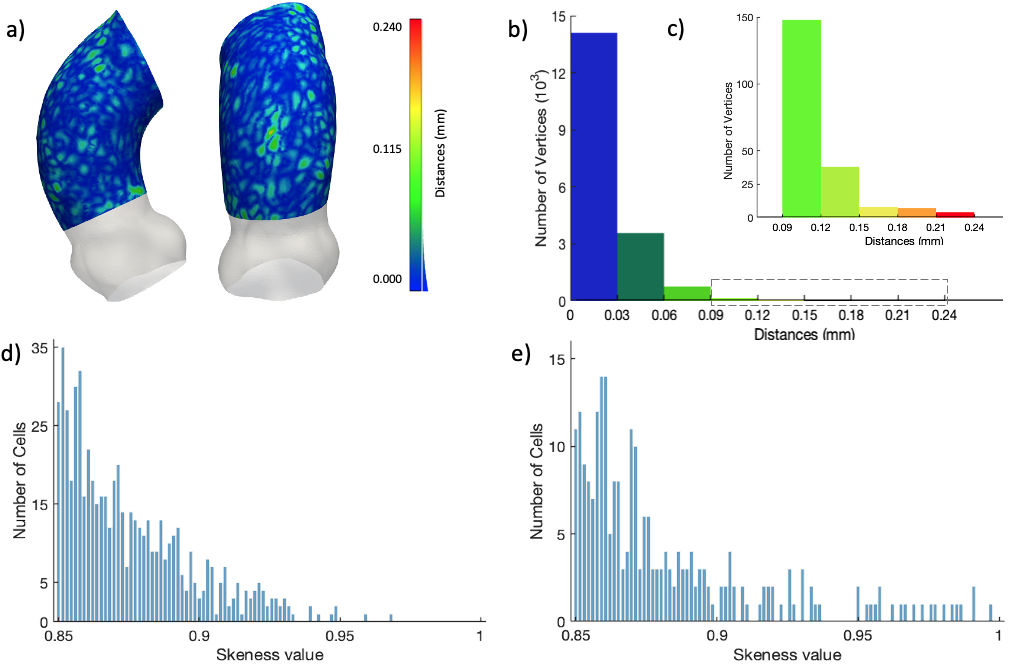}
    \caption{Results of RBF mesh morphing technique at geometry level: map of the distance between the the target 3D reconstructed phase at 20\% of cardiac cycle and the corresponding RBF solution (a); the bar-plot of nodes distance (b) and the detail of the columns with the lower number of nodes (c); cells with a skewness values \textgreater0.85 for the RBF solution corresponding to 40\% phase (d) and 80\% phase (e).}
    \label{fig:distance}
\end{figure}

Regarding the accuracy of the RBF mesh morphing technique, the assessment of geometrical effectiveness showed highest nodes distance for the 20\% phase of cardiac cycle with values ranging between 0 and 0.24 mm (Figure \ref{fig:distance}a-c). The maximum value of distance belonged to a very few number of mesh vertices (4 vertices), while the majority of nodes showed a value less than 0.03 mm.

The mesh quality analysis showed the introduction of cells with high skewness value (\textgreater0.85) as a results of the RBF mesh morphing. This condition was mainly observed at 40\% phase of cardiac cycle. Nevertheless only 19 cells exceeded the critical value of 0.95 together with an average value less than 0.3 at 80\% phase of cardiac cycle without problems for simulation convergence (Table \ref{tab:RBF}) \cite{fluent2009}. The distributions of skewness values for the two above mentioned cases are reported in Figure \ref{fig:distance}d-e.
However, these cells are confined in a small region corresponding to the superior boundary of the source points and no significant effects were reflected on the hemodynamic results.

\begin{table}[!ht]
\small\sf\centering
\begin{tabular}{l|ccc}
\toprule
 & $Avg$ & $\textgreater0.85$ & $\textgreater0.95$\\
\midrule
 $20\%$  & $0.268$ & $22$  & $0$\\
 $40\%$  & $0.274$ & $593$ & $2$\\
 $60\%$  & $0.276$ & $209$ & $2$\\
 $80\%$  & $0.269$ & $247$ & $19$\\
\bottomrule
\end{tabular}
  \vskip 0.2 cm
  \caption{Averaged values of skewness metric, number of cells with skewness values greater than 0.85 and greater than 0.95 (critical value for convergence) for RBF solutions for the models at selected phases of cardiac cycle.}
  \label{tab:RBF}
\end{table}  

Concerning the comparison of the fluid dynamic aspects between the CFD and CFD$_{RBF}$, differences were detected for the investigated variables. The evaluation of the through-plane velocity (Figure \ref{fig:piani}) revealed  differences in terms of both values and distribution. The wall movement increases the eccentricity of the flow along the vessel with an impingement on the flow versus the anterior artery wall. Our findings also show that this phenomenon is marked at time $T_3$ (Figure \ref{fig:NFE}a,e,i) while is reduced at $T_1$ (Figure \ref{fig:NFE}c,g,k).

Indeed, focusing on the morphological analysis of the $\Pi_2$ along the cardiac cycle, the maximum displacement of the centroid was equal to 50\% and the deformation of the maximum diameter was equal to 7.6\%. These variations detected by the analysis of the images of the datasets with CT gated were found to be responsible for the significant differences between the results of the standard CFD and the CFD$_{RBF}$ simulations.

The NFE index is widely studied in literature, in particular for aneurysmatic subjects for its impact on the ascending aorta in terms of WSS and TAWSS \cite{condemi2019}, and it is an important indicator of how the hemodynamics are largely led by geometrical aspects \cite{campobasso2018}. In term of hemodynamic indices, the differences between the CFD and the CFD$_{RBF}$ were addressed also by the LNH, OSI and TAWSS.
The role of LNH is to measure the local alignment between the velocity and the vorticity vector fields and it is associated with flow energy decay over time \cite{caro1996, Stonebridge1996}. 
In our study, a LNH threshold of $\pm$0.8 was chosen for visualization purposes to show the elevated helicity. The presence of helical flow in the thoracic aorta is based on the assumption that valve morphology and aorta curved geometry may promote its formation. Here we imposed a plug inlet condition for all the three models in order to stress out the influence of the morphological changes during the cardiac time. In fact, the differences observed in the helicity descriptors (Figure \ref{fig:LNH} and Figure \ref{fig:LNH_in_plane}) showed  
how the aortic helical flow structure firstly depends on the ascending aorta volume and even more markedly on the actual aorta deformation during cardiac cycle.

The main interesting findings of this work are related to the opposite sign of the predominant direction of rotation changes ($h_1$ and $h_3$) of the CFD$_{RBF}$ with respect to the CFD models.
On the other hand, the $h_2$ descriptor appears to be not affected by the ascending aorta variation along the cardiac cycle.
In particular, Figure \ref{fig:LNH}(g,h,i)
evidences that the topology of the flow begins to assume a double swirl-like structure, with an inversion in the helical rotation going from the inner (right handed, positive values of LNH) to the outer (left handed, negative values of LNH) aortic wall. This rearrangement of the flow starts from the ascending aorta and it involves the entire arch up to the supra-aortic vessels.The onset of bi-helical patterns and their dynamics can be appreciated in the second half of the systole up to the early diastole (Figure \ref{fig:LNH}h,i) where the flow with right-handed rotation is more desegregated.
This feature is more evident at T4 where the topological helical flow feature is more perturbed. During the diastole, the fluid rotational momentum dominates, resulting in counter-rotating helical patterns in the ascending aorta and proximal to a symmetric flow as systole goes (Figure \ref{fig:LNH_in_plane}g).
The evaluation of helical flow through LNH index was performed in several works starting from both \textit{in-vivo} \cite{morbiducci2011,garcia2017volumetric} and numerical simulations technique \cite{condemi2017,morbiducci2013inflow} mainly using CFD approach and no studies were performed with this specific comparison in mind. This different fluid rotation can be described by factors that, at this stage of investigation, are not easily identified.
We think that the observed flow structures are given by the combination of the superimposition of the axial flow at inlet and the ascending aorta wall motion; however, further investigation are required to confirm our preliminary results. 
Consistently to previous observations, differences  arise also from the evaluation of WSS indicators such as OSI (Figure \ref{fig:TAWSS_OSI}(a-c)) and TAWSS (Figure \ref{fig:TAWSS_OSI}(d-f)) mainly in terms of map distributions.
The representation of ascending aorta regions exposed to selected intervals of values (Figure \ref{fig:MapsPercentileOSI} and Figure \ref{fig:MapsPercentileTAWSS}) as well as the quantification of area exposed to disturbed shear (Table \ref{tab:Percentile}) further highlight the influence of different simulation strategies on OSI and TAWSS indicators.
The assumption on the shape of the phase at witch the geometry is considered affects the topology of the flow descriptors in the aorta. 
Focusing our attention on the 95th percentile value for OSI, an increase of about 14\% was obtained in case of CFD$_{40}$. The comparison between CFD$_{0}$ and CFD$_{40}$ shows an increase of 76\% in terms of area exposed to values below 5th percentile for TAWSS.
The results about a standard CFD and a CFD$_{RBF}$ confirm the importance of including the patient specific geometry variation during the cardiac cycle. 
The spread obtained in the values of WSS descriptors into the aorta differ both in intensity and topology. As an example for the disturbed shear, our results show that the standard CFD approach underestimates the surface areas in terms of OSI while overestimate TAWSS. For this reason, the incorporation of fully personalized  in vivo wall movement coupled with the inlet flow velocity could be a necessary step in the pipeline for subject-specific analysis of the aortic hemodynamics.
\\
There are several limitations that could weaken the findings of this study  and that could be considered in order to investigate in depth the added value of our approach.
Here we limited the morphing approach to the ascending aorta only. The reason for this choice is that the gated CT dataset was available only for this district. Due to the radiation dose necessary to obtain the whole aorta with a gated CT scan (mean radiation exposure of about 13.0 mGy \cite{satriano2018three} (5 mGy in our case), a dataset of entire thoracic aorta was not available.
However, the strategy applied here does not entail the generality of our findings because the ascending aorta is subject to the higher stress and strain values \cite{satriano2018three}. Moreover the movement of the ascending aorta is more complex, as a result of the motion from the beating heart \cite{jin2003effects}. However, the above limitation could be potentially overcome by using patient images from cine-MR. Despite the potentiality of the MR scan, it is worth to point out the difficulties to cope deformations of about or less of 1 mm with this methodology. Despite this limitations, MRI is recognised as an useful methodology to obtain patient-specific inlet boundary condition. According to previous works \cite{youssefi2018impact, condemi2019}, a further aspect to be taken into account is the application of 4D velocity map as inlet boundary condition to accurately simulate the real flow distribution at the inlet of ascending aorta. We plan to include this additional feature as a next step in order to replicate more faithfully the physical flow behaviour from both numerical \cite{Antonuccio2020} and experimental point of view \cite{celi20203d, vignali2019design}. 
A FSI simulation of the model was not performed and, apparently, this could seem to be the main limitation of this study. Subject-specific wall material properties were not available and, consequently, the adoption of values from literature would only add uncertainty to the results, posing a
great obstacle when performing patient-specific simulations. In this scenario, an addition uncertainty quantification would be necessary but this was not the aim of the study. 

A comparison between the computational requirements of the two simulation approaches shows similar results. This is because the RBF mesh morphing technique does not require an FE solver for the description of the vessel structure.
Segmentation, meshing, and setup of the CFD model take longer time, making the additional time required for including mesh morphing negligible. In terms of computational time, the CFD simulations are faster (3 times) with respect to CFD$_{RBF}$. It is worth to point out that the additional cost is not comparable to those required by a 2-way FSI simulation where the re-mesh process is called at each time step. Moreover, despite the time discrepancy, the CFD$_{RBF}$ can be considered affordable, as the physiological relevance increases considerably also considering that the mesh motion subroutine used in the simulation software are scripted and can be reused. Probably the routine be optimized even further, decreasing the additional computational effort even more to make more feasible the adoption of this technique in a clinical environment \cite{BiancoliniROM}.
In summary, a comprehensive evaluation of a novel RBF mesh morphing technique demonstrated that morphing with CT gated dataset can be applied to study patient specific hemodynamics.\\

\noindent Competing interests: None declared\\
Funding: None\\
Ethical approval: Not required


\bibliography{paper}

\begin{thebibliography}{10}
\expandafter\ifx\csname url\endcsname\relax
  \def\url#1{\texttt{#1}}\fi
\expandafter\ifx\csname urlprefix\endcsname\relax\def\urlprefix{URL }\fi
\expandafter\ifx\csname href\endcsname\relax
  \def\href#1#2{#2} \def\path#1{#1}\fi

\bibitem{Weigang2008}
E.~Weigang, F.~A. Kari, F.~Beyersdorf, M.~Luehr, C.~D. Etz, A.~Frydrychowicz,
  A.~Harloff, M.~Markl, {Flow-sensitive four-dimensional magnetic resonance
  imaging: flow patterns in ascending aortic aneurysms}, European Journal of
  Cardio-Thoracic Surgery 34~(1) (2008) 11--16.

\bibitem{den2010hemodynamic}
P.~M. den Reijer, D.~Sallee, P.~van~der Velden, E.~R. Zaaijer, W.~J. Parks,
  S.~Ramamurthy, T.~Q. Robbie, G.~Donati, C.~Lamphier, R.~P. Beekman, et~al.,
  Hemodynamic predictors of aortic dilatation in bicuspid aortic valve by
  velocity-encoded cardiovascular magnetic resonance, Journal of Cardiovascular
  Magnetic Resonance 12~(1) (2010) 4.

\bibitem{Qiao2011}
A.~Qiao, W.~fu, Y.-J. Liu, Study on hemodynamics in patient-specific thoracic
  aortic aneurysm, Theoretical and Applied Mechanics Letters 1 (01 2011).

\bibitem{harloff2010}
A.~Harloff, A.~Nu{\ss}baumer, S.~Bauer, A.~F. Stalder, A.~Frydrychowicz,
  C.~Weiller, J.~Hennig, M.~Markl, In vivo assessment of wall shear stress in
  the atherosclerotic aorta using flow-sensitive 4d mri, Magnetic Resonance in
  Medicine: An Official Journal of the International Society for Magnetic
  Resonance in Medicine 63~(6) (2010) 1529--1536.

\bibitem{taylor2009patient}
C.~A. Taylor, C.~Figueroa, Patient-specific modeling of cardiovascular
  mechanics, Annual review of biomedical engineering 11 (2009) 109--134.

\bibitem{celi2017multimodality}
S.~Celi, N.~Martini, L.~Emilio~Pastormerlo, V.~Positano, S.~Berti,
  Multimodality imaging for interventional cardiology, Current pharmaceutical
  design 23~(22) (2017) 3285--3300.

\bibitem{bakhshinejad2017}
A.~Bakhshinejad, A.~Baghaie, A.~Vali, D.~Saloner, V.~L. Rayz, R.~M. D’Souza,
  Merging computational fluid dynamics and 4d flow mri using proper orthogonal
  decomposition and ridge regression, Journal of biomechanics 58 (2017)
  162--173.

\bibitem{nayak2015}
K.~S. Nayak, J.-F. Nielsen, M.~A. Bernstein, M.~Markl, P.~D. Gatehouse, R.~M.
  Botnar, D.~Saloner, C.~Lorenz, H.~Wen, B.~S. Hu, et~al., Cardiovascular
  magnetic resonance phase contrast imaging, Journal of Cardiovascular Magnetic
  Resonance 17~(1) (2015) 71.

\bibitem{Petersson2012}
S.~Petersson, P.~Dyverfeldt, T.~Ebbers, Assessment of the accuracy of mri wall
  shear stress estimation using numerical simulations, Journal of magnetic
  resonance imaging : JMRI 36 (2012) 128--38.

\bibitem{Caballero2013}
A.~Caballero, S.~Lain, A review on computational fluid dynamics modelling in
  human thoracic aorta, Cardiovascular Engineering and Technology 4 (06 2013).

\bibitem{fanni2020}
B.~M. Fanni, K.~Capellini, M.~Di~Leonardo, A.~Clemente, E.~Cerone, S.~Berti,
  S.~Celi, Correlation between laa morphological features and computational
  fluid dynamics analysis for non-valvular atrial fibrillation patients,
  Applied Sciences 10~(4) (2020) 1448.

\bibitem{celi2014}
S.~Celi, S.~Berti, Three-dimensional sensitivity assessment of thoracic aortic
  aneurysm wall stress: a probabilistic finite-element study, European journal
  of cardio-thoracic surgery 45~(3) (2014) 467--475.

\bibitem{zhong2018}
L.~Zhong, J.-M. Zhang, B.~Su, R.~S. Tan, J.~C. Allen, G.~S. Kassab, Application
  of patient-specific computational fluid dynamics in coronary and
  intra-cardiac flow simulations: Challenges and opportunities, Frontiers in
  physiology 9 (2018) 742.

\bibitem{Bozzi2017}
S.~Bozzi, U.~Morbiducci, D.~Gallo, R.~Ponzini, G.~Rizzo, C.~Bignardi,
  G.~Passoni, Uncertainty propagation of phase contrast-mri derived inlet
  boundary conditions in computational hemodynamics models of thoracic aorta,
  Computer Methods in Biomechanics and Biomedical Engineering 20 (2017) 1--9.

\bibitem{condemi2017}
F.~Condemi, S.~Campisi, M.~Viallon, T.~Troalen, G.~Xuexin, A.~J. Barker,
  M.~Markl, P.~Croisille, O.~Trabelsi, C.~Cavinato, et~al., Fluid-and
  biomechanical analysis of ascending thoracic aorta aneurysm with concomitant
  aortic insufficiency, Annals of biomedical engineering 45~(12) (2017)
  2921--2932.

\bibitem{Taylor2010}
C.~Taylor, D.~Steinman, Image-based modeling of blood flow and vessel wall
  dynamics: Applications, methods and future directions, Annals of biomedical
  engineering 38 (2010) 1188--203.

\bibitem{avolio2013}
A.~Avolio, Arterial stiffness, Pulse 1~(1) (2013) 14--28.

\bibitem{Vulliemoz2002}
S.~Vulliémoz, N.~Stergiopulos, R.~Meuli, Estimation of local aortic elastic
  properties with mri, Magnetic Resonance in Medicine 47~(4) (2002) 649--654.

\bibitem{Jin2003}
S.~Jin, J.~Oshinski, D.~Giddens, Effects of wall motion and compliance on flow
  patterns in the ascending aorta, Journal of biomechanical engineering 125
  (2003) 347--54.

\bibitem{de2011}
L.~M. De~Heer, R.~P. Budde, P.~T.~M. Willem, A.~M. de~Vos, L.~A. van Herwerden,
  J.~Kluin, Aortic root dimension changes during systole and diastole:
  evaluation with ecg-gated multidetector row computed tomography, The
  international journal of cardiovascular imaging 27~(8) (2011) 1195--1204.

\bibitem{TSE2012}
K.~M. Tse, R.~Chang, H.~Lee, S.~Lim, S.~Venkatesh, P.~Ho, A computational fluid
  dynamics study on geometrical influence of the aorta on haemodynamics,
  European journal of cardio-thoracic surgery : official journal of the
  European Association for Cardio-thoracic Surgery 43 (07 2012).

\bibitem{trachet2015}
B.~Trachet, J.~Bols, J.~Degroote, B.~Verhegghe, N.~Stergiopulos,
  J.~Vierendeels, P.~Segers, An animal-specific fsi model of the abdominal
  aorta in anesthetized mice, Annals of biomedical engineering 43~(6) (2015)
  1298--1309.

\bibitem{pons2020fluid}
R.~Pons, A.~Guala, J.~Rodr{\'\i}guez-Palomares, J.~Cajas, L.~Dux-Santoy,
  G.~Teixid{\'o}-Tura, J.~Molins, M.~V{\'a}zquez, A.~Evangelista, J.~Martorell,
  Fluid--structure interaction simulations outperform computational fluid
  dynamics in the description of thoracic aorta haemodynamics and in the
  differentiation of progressive dilation in marfan syndrome patients, Royal
  Society Open Science 7~(2) (2020) 191752.

\bibitem{Mendez2018}
V.~Mendez, M.~Di~Giuseppe, S.~Pasta, Comparison of hemodynamic and structural
  indices of ascending thoracic aortic aneurysm as predicted by 2-way fsi, cfd
  rigid wall simulation and patient-specific displacement-based fea, Computers
  in Biology and Medicine 100 (2018).

\bibitem{campobasso2018}
R.~Campobasso, F.~Condemi, M.~Viallon, P.~Croisille, S.~Campisi, S.~Avril,
  Evaluation of peak wall stress in an ascending thoracic aortic aneurysm using
  fsi simulations: effects of aortic stiffness and peripheral resistance,
  Cardiovascular engineering and technology 9~(4) (2018) 707--722.

\bibitem{boccadifuoco2018impact}
A.~Boccadifuoco, A.~Mariotti, S.~Celi, N.~Martini, M.~Salvetti, Impact of
  uncertainties in outflow boundary conditions on the predictions of
  hemodynamic simulations of ascending thoracic aortic aneurysms, Computers \&
  Fluids 165 (2018) 96--115.

\bibitem{Fanni_2020a}
B.~M. Fanni, E.~Sauvage, S.~Celi, W.~Norman, E.~Vignali, L.~Landini,
  S.~Schievano, V.~Positano, C.~Capelli, A proof of concept of a non-invasive
  image-based material characterization method for enhanced patient-specific
  computational modeling, Cardiovascular Engineering and Technology (2020).
\newblock \href {https://doi.org/10.1007/s13239-020-00479-7}
  {\path{doi:10.1007/s13239-020-00479-7}}.

\bibitem{vignali2020modeling}
E.~Vignali, E.~Gasparotti, K.~Capellini, B.~M. Fanni, L.~Landini, V.~Positano,
  S.~Celi, Modeling biomechanical interaction between soft tissue and soft
  robotic instruments: importance of constitutive anisotropic hyperelastic
  formulations, The International Journal of Robotics Research (2020)
  0278364920927476.

\bibitem{avril2009vivo}
S.~Avril, J.~M. Huntley, R.~Cusack, In vivo measurements of blood viscosity and
  wall stiffness in the carotid using pc-mri, European Journal of Computational
  Mechanics/Revue Europ{\'e}enne de M{\'e}canique Num{\'e}rique 18~(1) (2009)
  9--20.

\bibitem{groth2019fast}
C.~Groth, U.~Cella, E.~Costa, M.~E. Biancolini, Fast high fidelity cfd/csm
  fluid structure interaction using rbf mesh morphing and modal superposition
  method, Aircraft Engineering and Aerospace Technology (2019).

\bibitem{evangelos2019radial}
M.~Evangelos~Biancolini, U.~Cella, Radial basis functions update of digital
  models on actual manufactured shapes, Journal of Computational and Nonlinear
  Dynamics 14~(2) (2019).

\bibitem{papoutsis2019aerodynamic}
E.~Papoutsis-Kiachagias, S.~Porziani, C.~Groth, M.~Biancolini, E.~Costa,
  K.~Giannakoglou, Aerodynamic optimization of car shapes using the continuous
  adjoint method and an rbf morpher, in: Advances in Evolutionary and
  Deterministic Methods for Design, Optimization and Control in Engineering and
  Sciences, Springer, 2019, pp. 173--187.

\bibitem{lantz2014}
J.~Lantz, P.~Dyverfeldt, T.~Ebbers, Improving blood flow simulations by
  incorporating measured subject-specific wall motion, Cardiovascular
  engineering and technology 5~(3) (2014) 261--269.

\bibitem{Porziani2017}
S.~Porziani, E.~Costa, M.~Biancolini, K.~Capellini, S.~Celi, Hemo-elastic study
  of ascending thoracic aorta aneurysms through rbf mesh morphing, 2017.

\bibitem{Capellini2018}
K.~Capellini, E.~Vignali, E.~Costa, E.~Gasparotti, M.~E. Biancolini,
  L.~Landini, V.~Positano, S.~Celi, {Computational Fluid Dynamic Study for aTAA
  Hemodynamics: An Integrated Image-Based and Radial Basis Functions Mesh
  Morphing Approach}, Journal of Biomechanical Engineering 140~(11) (2018).

\bibitem{groth2018medical}
C.~Groth, S.~Porziani, M.~Biancolini, E.~Costa, S.~Celi, K.~Capellini,
  M.~Rochette, V.~Morgenthaler, The medical digital twin assisted by reduced
  order models and mesh morphing, in: International CAE Conference, 2018.

\bibitem{biancolini2012}
M.~Biancolini, R.~Ponzini, L.~Antiga, U.~Morbiducci, A new workflow for patient
  specific image-based hemodynamics: parametric study of the carotid
  bifurcation, Computational Modelling of Objects Represented in Images III:
  Fundamentals, Methods and Applications. Rome, Italy (2012).

\bibitem{gallo2014virtual}
D.~Gallo, M.~E. Biancolini, R.~Ponzini, L.~Antiga, G.~Rizzo, A.~Audenino,
  U.~Morbiducci, A virtual test bench for hemodynamic evaluation of aortic
  cannulation in cardiopulmonary bypass, in: 11th world congress on
  computational mechanics. Barcelona, Spain, 2014.

\bibitem{Avril2020}
R.~Jayendiran, F.~Condemi, S.~Campisi, M.~Viallon, P.~Croisille, S.~Avril,
  Computational prediction of hemodynamical and biomechanical alterations
  induced by aneurysm dilatation in patient-specific ascending thoracic aortas,
  International Journal for Numerical Methods in Biomedical Engineering (2020)
  e3326.

\bibitem{boccadifuoco2016uncertainty}
A.~Boccadifuoco, A.~Mariotti, S.~Celi, N.~Martini, M.~V. Salvetti, Uncertainty
  quantification in numerical simulations of the flow in thoracic aortic
  aneurysms, Institute of Structural Analysis and Antiseismic Research, School
  of Civil Engineering, National Technical University of Athens, NTUA, Athens,
  Greece (2016) 6226--6249.

\bibitem{sigovan2011}
M.~Sigovan, M.~D. Hope, P.~Dyverfeldt, D.~Saloner, Comparison of
  four-dimensional flow parameters for quantification of flow eccentricity in
  the ascending aorta, Journal of Magnetic Resonance Imaging 34~(5) (2011)
  1226--1230.

\bibitem{shtilman1985}
L.~Shtilman, E.~Levich, S.~A. Orszag, R.~B. Pelz, A.~Tsinober, On the role of
  helicity in complex fluid flows, Physics Letters A 113~(1) (1985) 32--37.

\bibitem{morbiducci2009vivo}
U.~Morbiducci, R.~Ponzini, G.~Rizzo, M.~Cadioli, A.~Esposito, F.~De~Cobelli,
  A.~Del~Maschio, F.~M. Montevecchi, A.~Redaelli, In vivo quantification of
  helical blood flow in human aorta by time-resolved three-dimensional cine
  phase contrast magnetic resonance imaging, Annals of biomedical engineering
  37~(3) (2009) 516.

\bibitem{morbiducci2007helical}
U.~Morbiducci, R.~Ponzini, M.~Grigioni, A.~Redaelli, Helical flow as fluid
  dynamic signature for atherogenesis risk in aortocoronary bypass. a numeric
  study, Journal of biomechanics 40~(3) (2007) 519--534.

\bibitem{gallo2012}
D.~Gallo, D.~A. Steinman, P.~B. Bijari, U.~Morbiducci, Helical flow in carotid
  bifurcation as surrogate marker of exposure to disturbed shear, Journal of
  biomechanics 45~(14) (2012) 2398--2404.

\bibitem{malek1999}
A.~M. Malek, S.~L. Alper, S.~Izumo, Hemodynamic shear stress and its role in
  atherosclerosis, Jama 282~(21) (1999) 2035--2042.

\bibitem{fytanidis2014}
D.~Fytanidis, J.~Soulis, G.~Giannoglou, Patient-specific arterial system flow
  oscillation, Hippokratia 18~(2) (2014) 162.

\bibitem{lee2008geometry}
S.-W. Lee, L.~Antiga, J.~D. Spence, D.~A. Steinman, Geometry of the carotid
  bifurcation predicts its exposure to disturbed flow, Stroke 39~(8) (2008)
  2341--2347.

\bibitem{morbiducci2013inflow}
U.~Morbiducci, R.~Ponzini, D.~Gallo, C.~Bignardi, G.~Rizzo, Inflow boundary
  conditions for image-based computational hemodynamics: impact of idealized
  versus measured velocity profiles in the human aorta, Journal of biomechanics
  46~(1) (2013) 102--109.

\bibitem{leuprecht2002combined}
A.~Leuprecht, K.~Perktold, S.~Kozerke, P.~Boesiger, Combined cfd and mri study
  of blood flow in a human ascending aorta model, Biorheology 39~(3, 4) (2002)
  425--429.

\bibitem{romarowski2018patient}
R.~M. Romarowski, A.~Lefieux, S.~Morganti, A.~Veneziani, F.~Auricchio,
  Patient-specific cfd modelling in the thoracic aorta with pc-mri--based
  boundary conditions: A least-square three-element windkessel approach,
  International journal for numerical methods in biomedical engineering 34~(11)
  (2018) e3134.

\bibitem{zhu2018}
Y.~Zhu, R.~Chen, Y.-H. Juan, H.~Li, J.~Wang, Z.~Yu, H.~Liu, Clinical validation
  and assessment of aortic hemodynamics using computational fluid dynamics
  simulations from computed tomography angiography, Biomedical engineering
  online 17~(1) (2018) 53.

\bibitem{reymond2013}
P.~Reymond, P.~Crosetto, S.~Deparis, A.~Quarteroni, N.~Stergiopulos,
  Physiological simulation of blood flow in the aorta: comparison of
  hemodynamic indices as predicted by 3-d fsi, 3-d rigid wall and 1-d models,
  Medical engineering \& physics 35~(6) (2013) 784--791.

\bibitem{baumler2020}
K.~B{\"a}umler, V.~Vedula, A.~M. Sailer, J.~Seo, P.~Chiu, G.~Mistelbauer, F.~P.
  Chan, M.~P. Fischbein, A.~L. Marsden, D.~Fleischmann, Fluid--structure
  interaction simulations of patient-specific aortic dissection, Biomechanics
  and Modeling in Mechanobiology (2020) 1--22.

\bibitem{jin2003effects}
S.~Jin, J.~Oshinski, D.~P. Giddens, Effects of wall motion and compliance on
  flow patterns in the ascending aorta, J. Biomech. Eng. 125~(3) (2003)
  347--354.

\bibitem{singh2016effects}
S.~Singh, X.~Xu, J.~Pepper, C.~Izgi, T.~Treasure, R.~Mohiaddin, Effects of
  aortic root motion on wall stress in the marfan aorta before and after
  personalised aortic root support (pears) surgery, Journal of biomechanics
  49~(10) (2016) 2076--2084.

\bibitem{celi2012biomechanics}
S.~Celi, S.~Berti, Biomechanics and fe modelling of aneurysm: Review and
  advances in computational models, INTECH Open Access Publisher, 2012.

\bibitem{alimohammadi2015aortic}
M.~Alimohammadi, J.~M. Sherwood, M.~Karimpour, O.~Agu, S.~Balabani,
  V.~D{\'\i}az-Zuccarini, Aortic dissection simulation models for clinical
  support: fluid-structure interaction vs. rigid wall models, Biomedical
  engineering online 14~(1) (2015) 34.

\bibitem{boccadifuoco2018}
A.~Boccadifuoco, A.~Mariotti, K.~Capellini, S.~Celi, M.~V. Salvetti, Validation
  of numerical simulations of thoracic aorta hemodynamics: comparison with in
  vivo measurements and stochastic sensitivity analysis, Cardiovascular
  Engineering and Technology 9~(4) (2018) 688--706.

\bibitem{chandra2013fluid}
S.~Chandra, S.~S. Raut, A.~Jana, R.~W. Biederman, M.~Doyle, S.~C. Muluk, E.~A.
  Finol, Fluid-structure interaction modeling of abdominal aortic aneurysms:
  the impact of patient-specific inflow conditions and fluid/solid coupling,
  Journal of biomechanical engineering 135~(8) (2013).

\bibitem{fluent2009}
A.~Fluent, 12.0 theory guide, Ansys Inc 5~(5) (2009) 15.

\bibitem{condemi2019}
F.~Condemi, S.~Campisi, M.~Viallon, P.~Croisille, S.~Avril, Relationship
  between ascending thoracic aortic aneurysms hemodynamics and biomechanical
  properties, IEEE Transactions on Biomedical Engineering (2019).

\bibitem{caro1996}
C.~G. Caro, D.~J. Doorly, M.~Tarnawski, K.~T. Scott, Q.~Long, C.~L. Dumoulin,
  Non-planar curvature and branching of arteries and non-planar-type flow,
  Proceedings of the Royal Society of London. Series A: Mathematical, Physical
  and Engineering Sciences 452~(1944) (1996) 185--197.

\bibitem{Stonebridge1996}
P.~A. Stonebridge, P.~R. Hoskins, P.~Allan, J.~F.~F. Belch, {Spiral Laminar
  Flow in Vivo}, Clinical Science 91~(1) (1996) 17--21.

\bibitem{morbiducci2011}
U.~Morbiducci, R.~Ponzini, G.~Rizzo, M.~Cadioli, A.~Esposito, F.~M.
  Montevecchi, A.~Redaelli, Mechanistic insight into the physiological
  relevance of helical blood flow in the human aorta: an in vivo study,
  Biomechanics and modeling in mechanobiology 10~(3) (2011) 339--355.

\bibitem{garcia2017volumetric}
J.~Garcia, A.~J. Barker, J.~D. Collins, J.~C. Carr, M.~Markl, Volumetric
  quantification of absolute local normalized helicity in patients with
  bicuspid aortic valve and aortic dilatation, Magnetic resonance in medicine
  78~(2) (2017) 689--701.

\bibitem{satriano2018three}
A.~Satriano, Z.~Guenther, J.~A. White, N.~Merchant, E.~S. Di~Martino,
  F.~Al-Qoofi, C.~P. Lydell, N.~M. Fine, Three-dimensional thoracic aorta
  principal strain analysis from routine ecg-gated computerized tomography:
  feasibility in patients undergoing transcatheter aortic valve replacement,
  BMC cardiovascular disorders 18~(1) (2018) 76.

\bibitem{youssefi2018impact}
P.~Youssefi, A.~Gomez, C.~Arthurs, R.~Sharma, M.~Jahangiri,
  C.~Alberto~Figueroa, Impact of patient-specific inflow velocity profile on
  hemodynamics of the thoracic aorta, Journal of biomechanical engineering
  140~(1) (2018).

\bibitem{Antonuccio2020}
M.~N. Antonuccio, A.~Mariotti, S.~Celi, M.~V. Salvetti, Effects of the
  distribution in space of the velocity-inlet condition in hemodynamic
  simulations of the thoracic aorta, in: International Work-Conference on
  Bioinformatics and Biomedical Engineering, 2020, pp. 63--74.

\bibitem{celi20203d}
S.~Celi, E.~Gasparotti, K.~Capellini, E.~Vignali, B.~M. Fanni, L.~A. Ali,
  M.~Cantinotti, M.~Murzi, S.~Berti, G.~Santoro, et~al., 3d printing in modern
  cardiology, Current Pharmaceutical Design (2020).
\newblock \href {https://doi.org/10.2174/1381612826666200622132440}
  {\path{doi:10.2174/1381612826666200622132440}}.

\bibitem{vignali2019design}
E.~Vignali, Z.~Manigrasso, E.~Gasparotti, B.~Biffi, L.~Landini, V.~Positano,
  C.~Capelli, S.~Celi, Design, simulation, and fabrication of a
  three-dimensional printed pump mimicking the left ventricle motion, The
  International journal of artificial organs 42~(10) (2019) 539--547.

\bibitem{BiancoliniROM}
M.~E. Biancolini, K.~Capellini, E.~Costa, C.~Groth, M.~Rochette, S.~Celi, Fast
  interactive cfd evaluation of hemodynamics assisted by rbf mesh morphing and
  reduced order models: the case of ataa modelling, International Journal on
  Interactive Design and Manufacturing (2020).
\newblock \href {https://doi.org/10.1007/s12008-020-00694-5}
  {\path{doi:10.1007/s12008-020-00694-5}}.

\end{thebibliography}

\end{document}